\documentclass[dvips]{acta}
\usepackage{supertabular,lscape,epsfig}
\usepackage{amssymb}
\usepackage{amsmath}

\SetPages{0}{0}

\SetVol{58}{2008}

\begin{document}

\begin{Titlepage}

\Title{The Araucaria Project: The distance to the Small Magellanic 
Cloud from near infrared photometry of Type~II Cepheids\footnote{Based on
observations collected at the European Organisation for Astronomical
Research in the Southern Hemisphere, Chile with the ESO NTT for programmes
084.D-0591(B), 084.D-0591(E) and 084.D-0640(B)}}

\Author{A. Ciechanowska$^{1,2}$, G. Pietrzy\'nski$^{1,2}$, O. Szewczyk$^2$,
W. Gieren$^2$, and I. Soszy\'nski$^1$}
{$^1$Warsaw University Observatory, Al.~Ujazdowskie~4, 00-478 Warszawa,
Poland)\\
e-mail: (aciechan,pietrzyn,soszynsk)@astrouw.edu.pl\\
$^2$Universidad de Concepci\'on, Departamento de Fisica, Astronomy Group,
Casilla 160-C, Concepci\'on, Chile\\
e-mail: wgieren@astro-udec.cl}

\Received{Month Day, Year}
\end{Titlepage}

\Abstract{We have obtained deep near infrared $J$- and $K$-band observations of
14 BL~Herculis and 5 W~Virginis SMC stars from the OGLE~III survey
with the ESO New Technology Telescope equipped with the SOFI infrared
camera. From these observations, period-luminosity (P-L) relations in the 
$J$ and $K_s$ 2MASS bands were derived. The slopes of the 
$K$ and $J$ band relations of -2.15 $\pm$ 0.19 and -1.95 $\pm$ 0.24,
respectively, agree very well with the corresponding slopes derived  
previously for population~II Cepheids in globular clusters, Galactic 
bulge and in the Large Magellanic Cloud. The distance modulus to the SMC obtained 
from our data using
P-L relation derived for globular cluster Cepheids equals
$18.85 \pm 0.07\textrm{ (statistical)} \pm 0.07\textrm{ (systematic without 
including potential metallicity effect)}  $, which 
 within the uncertainties  agrees well with the 
results obtained  with other methods.}{Stars: Variables: W Virginis, BL Herculis, 
Galaxies: Distances and Redshifts, Galaxies: Individual: SMC, Galaxies: Stellar Content}

\section{Introduction}

The main goal of the long term program called the Araucaria project 
is to improve the calibration of the cosmic distance scale using extensive
and high quality photometric and spectroscopic
 observations of  several distance indicators in nearby galaxies
(Pietrzynski and Gieren 2009, Gieren et al. 2005). Comparative analysis 
of high quality 
optical and near-infrared photometry of major distance indicators like 
Cepheids, RR Lyrae stars, red clump stars and tip of the red giant branch 
(TRGB) magnitude, in nearby galaxies possesing very different environments
provide an opportunity to provide strong constraints on the population 
effects on these methods and construct a uniform set of accurate distances 
to the studied galaxies. Our results should also allow for a better calibration
of other techniques of distance determinations like Type~II Cepheids, blue 
supergiants, Anomalous Cepheids etc.

Type~II Cepheids are low-mass stars belonging to the disc and halo
populations. They have periods and amplitudes in a similar range as
classical Cepheids, but are about 1.5 - 2 mag fainter. We can divide them into
four different subgroups. Three of them: BL~Herculis stars, W~Virginis stars and
RV~Tauri stars show very similar period-luminosity relations and the
usual criterion of the division between them is the period. We adopted as 
in Soszy\'nski et al. (2010), 4 and 20 days as the period limits to distinguish 
these stars. The fourth group, peculiar W ~Virginis stars,
have light curves different from the typical light curves of  W~Vir stars 
and are  a little brighter (Soszy\'nski et al. 2008).         

Recent application of Type II Cepheids for distance determination to 
the Galactic Center (Kubiak and Udalski, 2003; Groenewegen, 
Udalski and Bono, 2008), and the LMC (Matsunaga et al. 2009, 
hereafter M09) confirmed that 
these stars  are very promising distance indicators. Therefore, an 
accurate calibration of their P-L relations, 
especially in the near infrared domain, where the extinction is 
very small, is of great importance. 

In this paper we will present the $J$ and $K_s$ band period-luminosity relations
based on 19 stars taken from the OGLE~III catalogue of type~II Cepheids in the SMC
(Soszy\'nski et al. 2010), compare it with similar relations for
objects in the LMC (M09), Galactic bulge (Groenewegen, Udalski and Bono, 
2008) and in globular clusters (Matsunaga et al. 2006, hereafter M06), and use 
it to calculate the SMC distance. 

\section{Observations}
We chose 14 BL~Her and 5 W~Vir stars from the
OGLE~III catalogue (Soszynski et al. 2010)
and observed them during three Araucaria observing runs, using the SOFI 
IR camera attached to  the ESO New Technology Telescope (NTT) at La Silla
Observatory. The wide-field mode was used to yield a 4.5 $\times$ 4.5
arcmin field of view and a scale of 0.288 arcsec per pixel. In order to
account for frequent sky-level variations in the near IR passbands a 
dithering technique was imployed. The detector readout time 
was adjusted to keep ADUs below 10000 (e.g. well within linearity regime of the
detector) during the observations. The whole sequence of dithered observations 
(typically 15-21 different positions) for a given field resulted in a 
total net integration time of 21 minutes in the
$K_s$-band (15 minutes for stars with period longer than 16 days) and 5.5
minutes in $J_s$, and allowed us to obtain a very good signal to noise ratio
of better than 50 for our target stars.
 
Detailed information about each observed field and observation conditions during
the exposures
is given in Table~1. (Note that name of each field is the name of the OGLE
star, on which this field is centered). 

\MakeTable{cccccccc}{12.5cm}{Observational information on the target fields}
{\hline \hline
Field name&R.A. 2000&Decl. 2000&Date of&HJD-2450000&HJD-2450000&Conditions&Extinction \\ 
 & & &observation&of $J_s$ exposure&of $K_s$ exposure& &$E(B-V)$ \\ \hline
OGLE-SMC-T2CEP-02 &00:34:53.51 &-72:58:45.9 &2010 Dec 02 &5168.707950 &5168.719090 &STD &0.087 \\
OGLE-SMC-T2CEP-03 &00:37:08.35 &-73:43:04.9 &2010 Dec 02 &5168.730445 &5168.741590 &STD &0.070 \\
OGLE-SMC-T2CEP-04 &00:38:20.36 &-73:17:16.4 &2010 Dec 04 &5170.618600 &5170.689680 &STD &0.070 \\
OGLE-SMC-T2CEP-05 &00:42:03.81 &-74:01:24.6 &2010 Dec 02 &5168.666730 &5168.677870 &STD &0.087 \\
OGLE-SMC-T2CEP-06 &00:42:16.01 &-73:39:13.5 &2010 Dec 03 &5169.554910 &5169.566020 &STD &0.078 \\
OGLE-SMC-T2CEP-08 &00:44:00.77 &-73:22:54.4 &2010 Dec 04 &5170.592745 &5170.603870 &STD &0.089 \\
OGLE-SMC-T2CEP-09 &00:44:12.37 &-72:59:28.5 &2010 Dec 27 &5193.702160 &5193.690865 &CLR &0.089 \\
OGLE-SMC-T2CEP-15 &00:49:36.92 &-73:10:01.4 &2010 Nov 05 &5141.677870 &5141.666550 &STD &0.101 \\
OGLE-SMC-T2CEP-16 &00:50:12.58 &-72:43:12.4 &2010 Dec 28 &5194.648940 &5194.637630 &CLR &0.101 \\
OGLE-SMC-T2CEP-17 &00:50:42.03 &-71:39:18.4 &2010 Dec 03 &5169.676400 &5169.687520 &STD &0.087 \\
OGLE-SMC-T2CEP-22 &00:54:46.72 &-73:48:32.6 &2010 Dec 03 &5169.740075 &5169.751170 &STD &0.087 \\
OGLE-SMC-T2CEP-27 &00:57:28.64 &-73:31:26.9 &2010 Dec 03 &5169.700485 &5169.711610 &STD &0.087 \\
OGLE-SMC-T2CEP-30 &00:57:40.76 &-73:03:04.9 &2010 Nov 06 &5142.690225 &5142.678905 &CLR &0.100 \\
OGLE-SMC-T2CEP-33 &00:59:03.09 &-72:28:32.2 &2010 Nov 06 &5142.655995 &5142.644685 &CLR &0.100 \\
OGLE-SMC-T2CEP-35 &01:00:35.01 &-73:46:57.9 &2010 Nov 07 &5143.576060 &5143.567220 &CLR &0.087 \\
OGLE-SMC-T2CEP-37 &01:03:46.50 &-74:07:28.8 &2010 Nov 07 &5143.682545 &5143.671235 &CLR &0.087 \\
 & & &2010 Dec 26 &5192.683115 &5192.671810 &STD & \\
OGLE-SMC-T2CEP-39 &01:06:40.91 &-73:07:05.0 &2010 Nov 07 &5143.606935 &5143.596055 &CLR &0.084 \\
 & & &2010 Dec 26 &5192.629215 &5192.617915 &STD & \\
OGLE-SMC-T2CEP-40 &01:08:46.85 &-71:51:09.4 &2010 Nov 07 &5143.754020 &5143.745155 &CLR &0.087 \\
 & & &2010 Dec 27 &5193.654555 &5193.645705 &CLR & \\
OGLE-SMC-T2CEP-42 &01:23:26.69 &-72:00:24.3 &2010 Dec 27 &5193.677110 &5193.665815 &CLR &0.087 \\ \hline
\multicolumn{8}{p{12.5cm}}{{\bf Note:} Extinction values are taken from
the reddening maps of Udalski et al. (1999). For objects located outside the
OGLE~II fields we adopted mean value of $E(B-V)$ from all OGLE~II fields (0.087 mag).
The entries STD and CLR mean photometric and clear observing conditions, respectively.}
}
 
\section{Data reductions and calibrations}
For all reductions and calibrations the pipeline developed in the course of
Araucaria Project was used. First, we applied a two-step sky level subtraction
process, including masking of stars with the IRAF\footnote{IRAF is distributed by the
National Optical Astronomy Observatories, which are operated by the   
Association of Universities for Research in Astronomy, Inc., under cooperative
agreement with the NSF.} xdimsum package
(Pietrzy\'nski and Gieren 2002). Next, each single image was flat-fielded
and stacked into the final deep field. Then point-spread function (PSF)
photometry, including aperture corrections, was performed in the same way
as described in Pietrzy\'nski et al. (2002).

The data collected under photometric conditions were calibrated onto 
the standard UKIRT
system using observations of 16 standard stars from the list published by 
Hawarden et al. (2001). The standard stars were observed
at different air masses and hour angles spread in between the regular 
target fields acquisition. The accuracy of the zero points
for both K and J band were estimated 
to be better than 0.02 mag. 
Next, the calibrated photometry was transformed onto the 2MASS photometric 
system using relations (37) and (39) given by Carpenter (2001). 
In order to perform an external check on the accuracy of the zero point of 
our photometry in each field we identified stars common to the 
2MASS Point Source Catalog (Wachter et al. 2003). In each case the zero point 
offset in both J and K bands was always smaller than 0.02 mag. 
The photometry of fields observed under
non-photometric conditions, were tied directly onto the 2MASS system
comparing our instrumental $j_s, k_s$ magnitudes with the 2MASS data. 

Since most of our stars have just single epoch measurements
we derived their mean magnitudes applying 
a simple method, following M09. 
Shortly, assuming that Type~II Cepheids have similar light curve shapes 
in $I$ and near infrared  $K$ and $J$ bands, and using  $I$-band light curves from (Soszy\'nski et al. 2010)
a correction between the observed random-phase K band magnitude and the mean magnitude
can be derived. Mean magnitudes obtained in this way will be 
called phase-corrected data.  

The calibrated data for all stars are presented in Table~2. In the columns
are star name (from OGLE~III catalogue), star type, period, $K_s$
magnitude and $J-K_s$ color (at the phase of measurement), and mean  $K_s$
magnitude and $J-K_s$ color. All errors are uncertainties as returned by {\sc   
Daophot}.

\MakeTable{ccccccc}{12.5cm}{Calibrated magnitudes and colors for SMC
population-II Cepheids}
{\hline \hline
Star ID& Type & Period & $K_s$ & $J-K_s$ & $<K_s>$ & $<J-K_s>$\\
(OGLE) & & (days) & (mag) & (mag) & (mag) & (mag)\\
\hline
OGLE-SMC-T2CEP-02 &BL Her & $1.3721870$ & $17.54 \pm 0.05$ & $0.41
\pm 0.06$ & $17.39 \pm 0.06$ & $0.40 \pm 0.08$ \\
OGLE-SMC-T2CEP-03 &W Vir & $4.3598789$ & $16.75 \pm 0.03$ & $0.50 \pm
0.04$ & $16.50 \pm 0.04$ & $0.54 \pm 0.05$ \\
OGLE-SMC-T2CEP-04 &W Vir & $6.5333997$ & $15.54 \pm 0.03$ & $0.19 \pm
0.04$ & $15.58 \pm 0.03$ & $0.16 \pm 0.04$ \\
OGLE-SMC-T2CEP-05 &W Vir & $8.2058890$ & $15.92 \pm 0.02$ & $0.55 \pm
0.03$ & $15.83 \pm 0.02$ & $0.57 \pm 0.03$ \\
OGLE-SMC-T2CEP-06 &BL Her & $1.2356136$ & $17.71 \pm 0.06$ & $0.42
\pm 0.07$ & $17.67 \pm 0.07$ & $0.41 \pm 0.09$ \\
OGLE-SMC-T2CEP-08 &BL Her & $1.4897859$ & $16.97 \pm 0.04$ & $0.26
\pm 0.04$ & $16.90 \pm 0.04$ & $0.27 \pm 0.05$ \\
OGLE-SMC-T2CEP-09 &BL Her & $2.9710719$ & $16.83 \pm 0.04$ & $0.44
\pm 0.04$ & $16.85 \pm 0.04$ & $0.47 \pm 0.04$ \\
OGLE-SMC-T2CEP-15 &BL Her & $2.5695964$ & $16.25 \pm 0.02$ & $0.11
\pm 0.03$ & $16.32 \pm 0.02$ & $0.11 \pm 0.03$ \\
OGLE-SMC-T2CEP-16 &BL Her & $2.1131980$ & $16.91 \pm 0.04$ & $0.37
\pm 0.04$ & $17.01 \pm 0.04$ & $0.39 \pm 0.05$ \\
OGLE-SMC-T2CEP-17 &BL Her & $1.2993097$ & $17.29 \pm 0.05$ & $0.29
\pm 0.06$ & $17.16 \pm 0.06$ & $0.31 \pm 0.07$ \\
OGLE-SMC-T2CEP-22 &BL Her & $1.4705261$ & $17.69 \pm 0.06$ & $0.29
\pm 0.07$ & $17.75 \pm 0.07$ & $0.29 \pm 0.09$ \\
OGLE-SMC-T2CEP-27 &BL Her & $1.5417249$ & $17.60 \pm 0.06$ & $0.27
\pm 0.07$ & $17.66 \pm 0.06$ & $0.21 \pm 0.08$ \\
OGLE-SMC-T2CEP-30 &BL Her & $3.3889388$ & $16.12 \pm 0.04$ & $0.52
\pm 0.05$ & $16.00 \pm 0.04$ & $0.52 \pm 0.05$ \\
OGLE-SMC-T2CEP-33 &BL Her & $1.8776865$ & $16.63 \pm 0.03$ & $0.38
\pm 0.03$ & $16.72 \pm 0.03$ & $0.31 \pm 0.04$ \\
OGLE-SMC-T2CEP-35 &W Vir & $17.1814841$ & $15.07 \pm 0.01$ & $0.36
\pm 0.01$ & $15.21 \pm 0.01$ & $0.36 \pm 0.02$ \\
OGLE-SMC-T2CEP-37 &BL Her & $1.5590709$ & $17.29 \pm 0.03^1$ & $0.25
 \pm 0.04^1$ & $17.18 \pm 0.03$ & $0.35 \pm 0.04$ \\
 & & & $17.06 \pm 0.03^2$ & $0.43 \pm 0.05^2$ & & \\
OGLE-SMC-T2CEP-39 &BL Her & $1.8875529$ & $16.95 \pm 0.03^1$ & $0.29
 \pm 0.04^1$ & $16.90 \pm 0.03$ & $0.33 \pm 0.04$ \\
 & & & $16.92 \pm 0.03^2$ & $0.38 \pm 0.05^2$ & & \\
OGLE-SMC-T2CEP-40 &W Vir & $16.1110373$ & $15.27 \pm 0.01^1$ & $0.53
 \pm 0.01^1$ & $14.93 \pm 0.01$ & $0.50 \pm 0.01$ \\
 & & & $15.41 \pm 0.01^3$ & $0.48 \pm 0.02^3$ & & \\
OGLE-SMC-T2CEP-42 &BL Her & $1.4874289$ & $17.29 \pm 0.04$ & $0.46
 \pm 0.04$ & $17.31 \pm 0.04$ & $0.50 \pm 0.06$ \\
\hline
\multicolumn{7}{p{9cm}}{$^1$ night 07-12-2009, $^2$ night 26-12-2009, $^3$ night 27-12-2009}
}
 
\section{Period-luminosity relations}
A fundamental issue  while applying P-L relations of 
pulsating  stars for distance measurement is to reassure the use 
of the correct fiducial period-luminosity relation. M06
calibrated such a relation for Type~II Cepheids using 46 stars
(3 with periods longer than 20 days and only 7 with shorter than 4 days) 
from 26 globular clusters and in the $K_s$ and $J$  bands obtained
\begin{equation}
M_{K_s}=-2.41(\pm 0.05)\log P -1.108(\pm 0.02)
\end{equation}
\begin{equation}
M_{J}=-2.23(\pm 0.05)\log P - 0.864 (\pm 0.03)
\end{equation}
with standard deviations of 0.14 and 0.16 mag, respectively.
In this paragraph we will check whether the  P-L 
relations of Matsunaga can be applied to the SMC Type~II 
Cepheids by computing the free least square fits for 
different subsamples of the observed target stars 
for both filters
and comparing the resulting slopes with the slope from equations (1) and (2). 

In order to produce reddening corrected 
P-L relations for the SMC Type~II Cepheids we adopted 
the E(B-V) values for all observed fields 
from the OGLE reddening maps (Udalski et al. 1999) as listed in Table 1, 
and  assume the reddening law
of Cardelli et al. (1989)  (e.g. $R_V=3.1$, $A_{K_s}=0.365 \times E(B-V)$, 
$A_{J}=0.866 \times E(B-V)$)
In the case of stars with
two  measurements we took mean value of them. 
Figures 1,2,3 and 4 show the
results for a mixed sample of BL~Her and W~Vir stars, and BL~Her stars solely,
respectively (unfortunately, we have not enough data to obtain a credible
relation for W~Vir stars alone). In all figures circled points and solid
line represent the phase-corrected data, whereas crosses and dashed lines are
for uncorrected observations.

\begin{figure}[htb]
\includegraphics{Fig1.epsi}
\FigCap{Reddening free $K_s$ band P-L relation of type~II Cepheids in the SMC. 
Filled circles and solid line represent phase-corrected data, while crosses and 
dashed line stand for uncorrected observations.}
\end{figure}

\begin{figure}[htb]
\includegraphics{Fig2.epsi}
\FigCap{Reddening free $K_s$ band P-L relation of BL~Her stars in SMC. Filled
circles and solid line represent phase-corrected data, crosses and dashed
line stand for uncorrected observations}
\end{figure}

\begin{figure}[htb]
\includegraphics{Fig3.epsi}
\FigCap{Reddening free $J$ band P-L relation of type~II Cepheids in the SMC.
Filled circles and solid line represent phase-corrected data, while crosses and
dashed line stand for uncorrected observations.
}
\end{figure}

\begin{figure}[htb]
\includegraphics{Fig3.epsi}
\FigCap{Reddening free $J$ band P-L relation of BL~Her stars in SMC. Filled
circles and solid line represent phase-corrected data, crosses and dashed
line stand for uncorrected observations.
}
\end{figure}

Last square solutions for BL~Her and W~Vir data yield:
\begin{eqnarray*}
K_s = -2.15(\pm 0.19) \log P + 17.59 (\pm 0.11), (\sigma = 0.29) \\
J = -2.05(\pm 0.23) \log P + 17.87 (\pm 0.13), (\sigma = 0.34) 
\end{eqnarray*} 
in the case of phase-corrected data and
\begin{eqnarray}
K_s = -2.05(\pm 0.20) \log P + 17.58 (\pm 0.11), (\sigma = 0.30) \nonumber \\
J = -1.95(\pm 0.24) \log P + 17.86 (\pm 0.13), (\sigma = 0.36) 
\end{eqnarray}
for uncorrected data.

For BL~Her data solely we obtained:
\begin{eqnarray*}
K_s = -3.04(\pm 0.61) \log P + 17.79 (\pm 0.17), (\sigma = 0.28) \\
J = -2.91(\pm 0.69) \log P + 18.06 (\pm 0.19), (\sigma = 0.33)
\end{eqnarray*} 
in the case of phase-corrected data and
\begin{eqnarray*}
K_s = -3.14(\pm 0.56) \log P + 17.82 (\pm 0.16), (\sigma = 0.26)\\
J = -3.02(\pm 0.66) \log P + 18.09 (\pm 0.19), (\sigma = 0.31)
\end{eqnarray*}
for uncorrected data.

As is clearly visible, the phase-correction method
did not allow to reduce the scatter on the observed $J$ and $K$-band 
P-L relations.  A method for  correction of 
mean magnitudes should take into account different shapes of the 
Type~II Cepheids light curves and the phase shift bewteen 
the maximum in the optical and near infrared bands.
Such a technique has been developed for Classical Cepheids
by Soszynski et al. (2005). Unfortunately, template light curves
in the near infrared bands are still lacking for Type~II Cepheids, so
in the following we decided to use the uncorrected, random-phase data. 

The scatter on the relations is significantly larger in comparison with the results 
obtained for Type II Cepheids in the LMC, the Galactic bulge and Galactic globular 
clusters (M09, M06, Groenewegen,Udalski and Bono, 2009). This fact is mostly related to 
the geometrical extension of the SMC in the line of sight, and also to using 
single-phase data
to construct the P-L relation. This, together with the relatively small range 
of periods of BL Her stars and the small number of W Virginis stars in our sample, 
prevented us from providing a strong test 
wether the $J$ and $K$-band P-L relations of these  stars and 
W Vir are co-linear in the SMC. 
Since within the uncertainties (2 $\sigma$) the is no evidence for 
a difference in the slopes 
derived for the combined set of Type~II Cepheids and BL Her stars alone,
and M09 have shown that the relations for BL Her and W Vir stars are co-linear 
in the LMC, we decided to use the combined sample for the distance determination
to the SMC.

\section{Distance determination}
The slopes of our $J$ and $K_s$-band  P-L relations obtained for the SMC Population II 
Cepheids ($K_s$: -2.05 $\pm$ 0.20, $J$: -1.95 $\pm$ 0.24) agree within their uncertainties 
with the corresponding slopes 
obtained for these stars in the LMC (M09; $K_s$: -2.278 $\pm$ 0.047, $J$: -2.163 $\pm$ 0.044), Galactic 
bulge (Groenewegen, Udalski and Bono, 2009; $K_s$: -2.24 $\pm$ 0.14), 
globular clusters (M06; $K_s$: -2.41 $\pm$ 0.05, $J$: 2.23 $\pm$ 0.05), and with the theoretical slopes 
(Di Criscienzo et al. 2007; $K_s$: -2.38 $\pm$ 0.02, $J$: -2.29 $\pm$ 0.04). 

Therefore, we used the calibrations of 
M06 as a fiducial P-L relations to calculate distance to the SMC based on 
our data. The least square fitting gives the following  SMC distance moduli:
\begin{equation}
{(m-M)}_{0}  = 18.85 \pm 0.07 \textrm{(statistical) ($K_s$ band)}
\end{equation}
\begin{equation}
{(m-M)}_{0}  = 18.85 \pm 0.09 \textrm{(statistical) ($J$ band)}
\end{equation}

These results are (by coincidence) identical to each other and 
 to the distance modulus of  $18.85 \pm 0.11$ 
derived from OGLE optical photometry of Type~II Cepheids (Majaess et al. 2009). 
Within its uncertainties they also agree with the SMC distance moduli obtained from
 near-infrared photometry of RR~Lyr variables (Szewczyk et al. 2009) and red
clump stars (Pietrzy\'nski  al. 2003). Moreover the differencial distances 
Galactic bulge - LMC - SMC calculated from the $K$  band photometry 
of Type~II Cepheids agree very well with the corresponding data obtained for 
other distance indicators (Cepheids, TRGB, RR Lyrae, etc;  e.g. Udalski 2000, 
Groenewegen, Udalski, Bono 2009).
Our results suggest that any population
dependence of the slope and zero point of the $K_s$ and $J$ band P-L relations 
for  Type~II Cepheids should be small. In line with this conclusion, M06 
found also no evidence for significant population effects on JHK band 
P-L relatios of Type II Cepheids from the globular cluster data they used.
These conclusions are further
 supported by theoretical models  (Bono, Caputo \& Santolamazza, 1997, Di Criscienzo 
et al. 2007), which suggest that the properties of Type~II Cepheids 
should be only minimally affected by metallicity.

In order to estimate the systematic error on the SMC distance determined in this paper
we took into account the 
errors associated with photometric zero points (0.02 for both filters), transformation 
onto the 2MASS system (0.01), reddening (0.01 for $K_s$ and 0.02 for $J$ band)
 and the adopted fiducial P-L relation of M06. 
M06 combined the sample of Type~II Cepheids observed in different clusters 
using distance moduli calculated based on the magnitudes of the horizontal branches 
of the clusters. They adopted the relation of Gratton (2003) 
(${\rm {M}_{V}}$=0.22[Fe/H]+0.89), calibrated based on the main sequence distances. 
Therefore it is rather difficult to estimate a realistic systematic error of 
the zero point of the M06 calibration. Independently, the zero point of 
the P-L relation for Type~II Cepheids may be calibrated using two Galactic stars with 
pulsational parallaxes (Feast et al. 2008) with a precision of 0.06 mag.
If we adopt both zero point calibrations and calculate a distance to the LMC
the difference in the corresponding distance moduli is  0.04 mag only. 
Therefore it is reasonable to assume 0.06 as a realistic error of the zero 
point of the current calibrations of the P-L relation of Type~II 
Cepheids. Combining all errors quadratically we obtain 0.07 mag 
as the total systematic error of the SMC distance. This error does not 
contain the possible effect of a metallicty dependence on the 
slope and zero point of the Type II Cepheids P-L relation. 
However, as has been discussed above such an effect is supposed 
to be small.

\section{Summary and conclusions}
We presented precision $J$- and $K_s$-band photometry of 19 Type~II Cepheids in the
SMC. The slope of the resulting $J$ and  $K_s$ band P-L relations were
found to be, within the errors, in good agreement with the corresponding slopes 
obtained for Type~II Cepheids in the LMC (M09), Galactic bulge (Groenewegen, 
Udalski and Bono) and  globular clusters (M06). 

Assuming the fiducial P-L relation of M06 the following SMC distance modulus
was derived from our data:

\begin{equation}
{(m-M)}_{0} = 18.85 \pm 0.07 \textrm{(statistical)} \pm 0.07 \textrm{(systematic)}
\end{equation}

This result agrees very well with recent distance determinations to the SMC
based on other techniques. Moreover, the differencial 
distances between the LMC, SMC and Galactic Bulge derived from Type II
Cepheids are very similar to those obtained from Cepheids, TRGB, RR Lyrae stars and 
red clump stars, which suggests that the population effects on the slope and/or 
zero point of the Type~II Cepheid P-L relation must be small. 
Our results confirm that  Type~II Cepheids (W Vir and BL Her)
are very good standard candles. 

\Acknow{We gratefully acknowledge financial support for this
work from the Chilean Center for Astrophysics FONDAP 15010003, and from
the BASAL Centro de Astrofisica y Tecnologias Afines (CATA) PFB-06/2007.
Support from the Polish grants N203 387337 and N203 509938,
 and the FOCUS
subsidy of the Fundation for Polish Science (FNP) is also acknowledged.
}

\end{document}